\documentstyle[12pt,amsfonts]{article}

\newcommand{\mbf}[1]{\mbox{\boldmath$ #1$}}

\newcommand{\be}{\begin{equation}}
\newcommand{\ee}{\end{equation}}
\newcommand{\ba}{\begin{eqnarray}}
\newcommand{\ea}{\end{eqnarray}}

\begin{document}

\begin{center}
{{\LARGE {\bf Electromagnetic Pulse Propagation in Passive Media by the
Lanczos Method}}}

\vskip 1cm

{Andrei G. BORISOV ${}^{a,c,1}$ \ \ and \ \ Sergei V. SHABANOV ${}^{b,c,2}$}

\vskip 1cm

{\it ${}^a$\ Laboratoire des Collisions Atomiques et Mol\'eculaires,\\
UMR CNRS-Universit\'{e} Paris-Sud 8625, B\^{a}t. 351,\\
Universit\'{e} Paris-Sud, 91405 Orsay CEDEX, France\\
${}^b$\ Department of Mathematics, University of Florida, Gainesville,\\
FL 32611, USA\\
${}^c$\ Donostia International Physics Center (DIPC), P. Manuel de\\
Lardizabal 4, 20018 San Sebasti\'{a}n, Spain }
\end{center}

\begin{abstract}
Maxwell's equations are cast in the form of the Schr\"{o}dinger equation.
The Lanczos propagation method is used in combination with the fast Fourier
pseudospectral method to solve the initial value problem. As a result, a
time-domain, unconditionally stable, and highly efficient numerical
algorithm is obtained for the propagation and scattering of broad-band
electromagnetic pulses in dispersive and absorbing media. As compared to
conventional finite-difference time-domain methods, an important advantage
of the proposed algorithm is a dynamical control of accuracy: Variable time
steps or variable computational costs per time step with error control are
possible. The method is illustrated with numerical simulations of
extraordinary transmission and reflection in metal and dielectric gratings
with rectangular and cylindrical geometry.
\end{abstract}

\begin{description}
\item  Keywords: Lanczos algorithm; Maxwell's equations; time-domain
algorithms; pseudospectral methods; gratings

\item  $^{1}$ borisov@lcam.u-psud.fr

\item  $^{2}$ shabanov@phys.ufl.edu
\end{description}

\newpage

\section{Introduction}

Recent developments in photonics and nanostructure materials \cite
{Photonics,PhotonicsGr} have increased interest in efficient and accurate
algorithms for numerical simulations of the propagation and scattering of
short (broad band) laser pulses in generic passive (dispersive and
absorbing) media. Time-domain approaches for solving the Maxwell's equations
might be more suitable for this purpose than frequency domain methods
because the sought-for information, e.g., the scattering matrix, can be
obtained within a desired frequency range by a single propagation. Coupled
with laser ellipsometry of broad band pulses, fast simulations of expected
resonance patterns in the scattering amplitude appear to be an efficient
tool to control quality of manufactured photonic devices. Unconditionally
stable algorithms are especially advantageous for such tasks because of
their applicability to practically all materials and geometries without any
assessment of admissible values of the system parameters. Another attractive
property of time domain methods is their universality. The very same
algorithms can be used to calculate static properties of the system (e.g., a
band structure of photonic crystals), to simulate the electromagnetic pulse
propagation in non-linear materials as well as in media with time-dependent
properties.

The advantages of time-domain methods have been for a long time recognized
in quantum mechanics where they are extensively used in the fields of
chemical reaction dynamics\cite{Chemistry}, laser - matter
interactions\thinspace \cite{LaserMatter}, etc. Highly efficient and
accurate tools have been developed for the wave packet propagation and
analysis of the results \cite{Chemistry,Review,Ps,FilterDi,Flux,Smat}. Since
Maxwell's equations can be cast in the form of the Schr\"{o}dinger equation,
it is then natural to extend time-domain methods of quantum mechanics to
numerical electrodynamics. Some realizations of this idea are rooted to the
path integral representation of quantum theory (the Lie-Trotter product
formula \cite{nelson} or the split operator method \cite{Review,feit,Split}%
). The others exploit polynomial approximations of the fundamental solution
of the Sch\"{o}dinger equation. For instance, the Chebychev time-propagation
technique has been recently used to simulate the electromagnetic pulse
propagation in {\it non-absorbing} media \cite{Kole,Neuhauser}.

Here it is proposed to use the Lanczos algorithm \cite{lanczos} to obtain an
unconditionally stable, time-domain solver of Maxwell's equations for
passive media. The method allows for a dynamical control of accuracy,
meaning that computational costs are constantly optimized in due course of
simulations with error control. In brief, the approach can be summarized as
follows. Maxwell's equations are written in the form of the Schr\"{o}dinger
equation which is then solved by the Lanczos propagation scheme \cite
{park,Review} (Section II). The difference with the well studied quantum
mechanical case is that the wave function is a multi-dimensional vector
field and the Hamiltonian is non-Hermitian for absorbing media. The split
operator method \cite{Review,feit,Split} has been used to include
attenuation into the Lanczos propagation scheme, while preserving its
unconditional stability (Section III). The action of the Hamiltonian on the
wave function is computed by means of the Fourier pseudospectral method
introduced in \cite{kosloff}.

The accuracy of the method is investigated in Section IV. In Section V the
Lanczos propagation scheme is applied to various gratings. In particular, a
resonant extraordinary reflection of a periodic array of parallel dielectric
cylinders is observed. This effect is similar to the Wood anomalies \cite
{Wood} and related to the existence of stationary (trapped) electromagnetic
waves with wave vectors parallel to the discrete translation symmetry axis
of the system. Simulations of the scattering of broad band pulses on
metallic grating and grooves, whose dielectric properties are described by
the Drude model, are performed to demonstrate that the Lanczos propagation
scheme is able to reproduce the results known in the literature and obtained
by different means (by finite differencing schemes or by the scattering
matrix method).

\section{The Lanczos method for Maxwell's equations}

\setcounter{equation}0

Consider first the case of non-dispersive media. Let ${\bf D}$ and ${\bf B}$
be electric and magnetic inductions, respectively, and ${\bf E}$ and ${\bf H}
$ the corresponding fields so that ${\bf D}=\varepsilon {\bf E}$ and ${\bf B}%
=\mu {\bf H}$ where $\varepsilon $ and $\mu $ are positive, symmetric,
position dependent matrices for generic non-isotropic and non-homogeneous
media. For isotropic media, $\varepsilon $ and $\mu $ are scalars. At
interfaces of different media, the boundary conditions are enforced
dynamically, that is, $\varepsilon $ and $\mu $ are allowed to have
discontinuities. Maxwell's equations are rewritten as: 
\begin{equation}
i\dot{\psi}={\cal H}\psi \ ,\ \ \ \ \psi =\left( 
\begin{array}{c}
{\bf E} \\ 
{\bf H}
\end{array}
\right) \ ,\ \ \ \ {\cal H}=\left( 
\begin{array}{cc}
0 & ic\varepsilon ^{-1}{\mbox{\boldmath$ \nabla$}\times } \\ 
-ic\mu ^{-1}{\mbox{\boldmath$ \nabla$}\times } & 0
\end{array}
\right) \ .  \label{2.1}
\end{equation}
The over-dot denotes the time derivative, and $c$ is the speed of light in
the vacuum. One can also use the electromagnetic inductions as independent
variables instead of the fields. The Hamiltonian ${\cal H}$ must then be
modified accordingly. The initial value problem is solved by applying the
evolution operator (or the fundamental solution) to the initial wave
function 
\begin{equation}
\psi (t)=e^{-i{\cal H}t}\psi (0)\ .  \label{2.2}
\end{equation}
The Hamiltonian is a Hermitian operator, ${\cal H}^{\dagger }={\cal H}$,
with respect to the measure scalar product 
\begin{equation}
(\psi _{1},\psi _{2})=\int ({\bf D}_{1}\cdot {\bf E_{2}}+{\bf B}_{1}\cdot 
{\bf H}_{2})\,d{\bf r}\equiv \int \psi _{1}^{\dagger }\kappa \psi _{2}\,d%
{\bf r}\ .  \label{2.3}
\end{equation}
The symmetric positive matrix $\kappa $ is block-diagonal, with the blocks
being $\varepsilon $ and $\mu $. The norm of the wave function with respect
to the scalar product (\ref{2.3}) is proportional to the electromagnetic
energy and is conserved because the evolution operator is unitary.

In numerical simulations, the Hilbert space is projected onto a finite
dimensional Euclidean space so that $\psi $ becomes a vector whose
components are values of the wave function at sites of a finite spatial
grid. In the grid representation, ${\cal H}$ is a matrix. If the Hamiltonian
is Hermitian, it is then convenient to have ${\cal H}$ as an explicitly
Hermitian matrix. In the Maxwell theory, this can be achieved if, before
projecting onto the grid, the wave function and the Hamiltonian are scaled 
\begin{equation}
\psi \rightarrow \kappa ^{-1/2}\psi \ ,\ \ \ \ {\cal H}\rightarrow \kappa
^{-1/2}{\cal H}\kappa ^{1/2}\ .
\end{equation}
In the representation (\ref{2.1}) we have 
\begin{equation}
{\bf E}\rightarrow \varepsilon ^{-1/2}{\bf E},\ {\bf H}\rightarrow \mu
^{-1/2}{\bf H},\ \ {\cal H}\rightarrow \left( 
\begin{array}{cc}
0 & ic\varepsilon ^{-1/2}{\mbox{\boldmath$ \nabla$}\times }\mu ^{-1/2} \\ 
-ic\mu ^{-1/2}{\mbox{\boldmath$ \nabla$}\times }\varepsilon ^{-1/2} & 0
\end{array}
\right) \ .
\end{equation}
The scaled Hamiltonian is Hermitian with respect to the conventional scalar
product in the space of square integrable functions, and, hence, it is a
Hermitian matrix, when projected onto the grid. The action of spatial
derivatives is calculated within the pseudospectral approach based on the
Fourier grid representation of the wavefunction and the fast Fourier
transform. In what follows, only consecutive actions of ${\cal H}$ on wave
functions are required.

A direct use of (\ref{2.2}) implies a diagonalization of ${\cal H}$, which
is not feasible if the matrix size is too large. Various numerical
approximations are based on the semigroup property of the evolution operator 
\begin{equation}
\psi (t+\Delta t)=e^{-i\Delta t{\cal H}}\psi (t)\ .  \label{2.4}
\end{equation}
In a local propagation scheme the exponential can be approximated by a
polynomial for a sufficiently small time step $\Delta t$. The basic idea of
the Lanczos propagation method is that the exact solution $\psi (t+\Delta t)$
is projected onto the Krylov subspace associated with the initial state $%
\psi (t)$ and the Hamiltonian, $\psi (t+\Delta t)\rightarrow \psi
^{(n)}(t+\Delta t)\equiv {\cal P}_{n}\psi (t+\Delta t)\in {\sf K}_{n}$,
where ${\cal P}_{n}^{\dagger }{\cal =P}_{n}$, ${\cal P}_{n}^{2}{\cal =P}_{n}$%
, and 
\[
{\sf K}_{n}={\rm Span}\,\left( \psi (t),{\cal H}\psi (t),...,{\cal H}%
^{n-1}\psi (t)\right) \ .
\]
The accuracy of such an approximation is $O(\Delta t^{n})$. The Hamiltonian
is projected accordingly, ${\cal H}\rightarrow {\cal H}^{(n)}\equiv {\cal P}%
_{n}{\cal H}{\cal P}_{n}$. Thus, 
\begin{equation}
\psi (t+\Delta t)\approx \psi ^{(n)}(t+\Delta t)=e^{-i\Delta t{\cal H}%
^{(n)}}\psi ^{(n)}(t)\ .  \label{2.8}
\end{equation}
The projection is done via an orthonormal basis for ${\sf K}_{n}$ which is
constructed by means of the Lanczos recursion algorithm \cite{lanczos}. In
this basis, the matrix ${\cal H}^{(n)}$ is Hermitian and tridiagonal.
Typically, just a few orders are sufficient ($n\leq 9$) so that $n$ is much
smaller than the dimension of ${\cal H}$ and the matrix ${\cal H}^{(n)}$ can
easily be diagonalized. The dimension $n$ may be set differently at each
time step, depending on the current vector $\psi (t)$, and is determined by
a pre-set required accuracy. In particular, it allows to avoid excessive
actions of ${\cal H}$ on the wave function. This feature leads to a
dynamical optimization of computational costs with error control, which is
one the greatest advantages of the Lanczos method.

A detailed discussion of the Lanczos recursion algorithm and its application
to the wave packet propagation can be found elsewhere \cite{park,Review}.
Here only a brief summary is given with notations used later in the text.
Let $\psi _{0}=\psi (t)$ where $\psi (t)$ is assumed to be normalized so
that $\Vert \psi _{0}\Vert =1$. Due to the linearity of the Schr\"{o}dinger
equation one can always scale $\psi _{0}$ by a number and rescale it back
after applying the infinitesimal evolution operator. Define 
\begin{eqnarray}
\alpha _{0} &=&(\psi _{0},{\cal H}\psi _{0})\ ,  \label{2.5c} \\
\phi _{1} &=&({\cal H}-\alpha _{0})\psi _{0}\ , \\
\psi _{1} &=&\phi _{1}/\Vert \phi _{1}\Vert \ .
\end{eqnarray}
By construction, $\psi _{1}$ and $\psi _{0}$ are orthonormal. For $%
k=2,3,...,n-1$ the rest of the basis for ${\sf K}_{n}$ is generated by the
recursion relation 
\begin{eqnarray}
\alpha _{k-1} &=&(\psi _{k-1},{\cal H}\psi _{k-1})\ ,  \label{2.7} \\
\beta _{k-2} &=&(\psi _{k-2},{\cal H}\psi _{k-1})\ , \\
\phi _{k} &=&({\cal H}-\alpha _{k-1})\psi _{k-1}-\beta _{k-2}\psi _{k-2}\ ,
\\
\psi _{k} &=&\phi _{k}/\Vert \phi _{k}\Vert \ .
\end{eqnarray}
By construction, the vector ${\cal H}\psi _{j}$ is a linear combination of $%
\psi _{j-1}$, $\psi _{j}$, and $\psi _{j+1}$. Hence, in the Lanczos basis
the matrix ${\cal H}_{ij}^{(n)}=\left( \psi _{i},{\cal H}\psi _{j}\right) $
is tridiagonal. Elementary calculations show that the diagonal elements are $%
{\cal H}_{jj}^{(n)}=\alpha _{j}=\bar{\alpha}_{j}$, the upper and lower
superdiagonals are ${\cal H}_{jj+1}^{(n)}={\cal H}_{j-1j}^{(n)}=\beta _{j}=%
\bar{\beta}_{j}$.

Let $U$ be a unitary transformation such that $U^{\dagger }{\cal H}^{(n)}U$
is a diagonal matrix, and $E_{j}$ be eigenvalues of ${\cal H}^{(n)}$. The
approximate solution (\ref{2.8}) is obtained by expanding the wave function
over the Lanczos basis and solving the Sch\"{o}dinger equation for the
expansion coefficients: 
\begin{equation}
\psi ^{(n)}(t+\Delta t)=\sum_{k,j=0}^{n-1}\bar{U}_{jk}\,e^{-i\Delta
tE_{j}}\,U_{j0}\,\psi _{k}\equiv \sum_{k=0}^{n-1}c_{k}(\Delta t)\psi _{k}\ ,
\label{2.10}
\end{equation}
where the initial condition $c_{k}(0)=\delta _{k0}$ has been taken into
account. Since ${\cal H}^{(n)}$ is Hermitian, the evolution preserves the
norm 
\begin{equation}
\Vert \psi ^{(n)}(t+\Delta t)\Vert ^{2}=\Vert \psi ^{(n)}(t)\Vert ^{2}=\Vert
\psi _{0}\Vert ^{2}=1\ .  \label{2.11}
\end{equation}
Thus, the algorithm is unconditionally stable because the norm of the
amplification matrix ${\cal G}^{(n)}(\Delta t)$, defined by $\psi
^{(n)}(t+\Delta t)={\cal G}^{(n)}(\Delta t)\psi (t)$, is uniformly bounded, $%
\Vert {\cal G}^{(n)}(t)\Vert \leq 1$, for all parameters of the Hamiltonian
and $\Delta t\geq 0$.

The accuracy of the algorithm can be estimated from the following
observation \cite{park}. The norm of a projection of the exact solution onto
the orthogonal complement of ${\sf K}_{n}$ can be used as a measure of
accuracy of the Lanczos algorithm. By expanding the exponential in the right
hand side of (\ref{2.4}) into the Taylor series, it is clear that the
contribution of the term $(\Delta t{\cal H})^{n+1}\psi (t)$, which has no
projection onto ${\sf K}_{n}$, can only be captured by the approximate
solution if the larger Krylov space ${\cal K}_{n+2}$ is used in the Lanczos
algorithm, which, in turn, implies that the vector $c_{j}(\Delta t)$
acquires two additional components. Thus, the accuracy of the Lanczos
algorithm can be controlled, for example, by demanding that the absolute
value of $c_{n-1}(\Delta t)$ is less than a specified small number $\epsilon 
$. Note that $|c_{n-1}(\Delta t)|\sim O(\Delta t^{n-2})$ as one can deduce
from (\ref{2.10}) and the tridiagonal structure of ${\cal H}^{(n)}$ in the
Lanzcos basis. To ensure that the norm of the projection of $\psi (t+\Delta
t)$ onto the orthogonal complement of ${\sf K}_{n}$ is small, we demand that 
\begin{equation}
|c_{n-3}(\Delta t)|^{2}+|c_{n-2}(\Delta t)|^{2}+|c_{n-1}(\Delta t)|^{2}\leq
\epsilon   \label{eps}
\end{equation}
where $\epsilon \sim 10^{-14}$ in our calculations. To satisfy (\ref
{eps}), the time step $\Delta t$, or the dimension of the Krylov subspace $n$%
, or both can be varied to minimize computational costs. This is the
aforementioned dynamical control of accuracy in the Lanczos propagation
method. In our simulations, $\Delta t$ has been kept fixed, while (\ref{eps}%
) has been used to determine a minimal $n$ for each time step.

\section{Including attenuation by the split method}

\setcounter{equation}0

The response function of a passive medium in an applied electromagnetic
field must satisfy the causality condition. A common way to model the causal
response function is to assume that the medium polarization and
magnetization satisfy a linear differential equation in time in which a
non-homogeneous term is proportional to the applied field (for linear
media). The Maxwell's equations in passive media appear then to be a system
of (high-order) differential equations to which numerical algorithms are
applied \cite{DiffEquationForResponce,Taf}. Any\ system of high-order
differential equations can be converted into a system of first-order
differential equations by introducing auxiliary dynamical variables. This
idea is used to convert Maxwell's equations for passive media into the
Schr\"{o}dinger equation (\ref{2.1}) in which the wave function contains
additional components that describe dynamics of the medium polarization and
magnetization. Due to absorption the time evolution is no longer unitary.

It must be noted that absorption of the wave packet is required in numerical
simulations of scattering problems in which the pulse shape is to be
computed in the asymptotic region. Indeed, when the front edge of the pulse
reaches the grid boundary, it will be reflected or re-appear on the other
side of the grid, depending on the boundary conditions. To avoid an
artificial interference of the scattered pulse with itself, a layer of an
absorbing medium is necessary at the grid boundary \cite{CAPS}.

Here a simple way is proposed to include the attenuation of the wave packet
amplitude into the Lanczos method, while maintaining the unconditional
stability of the algorithm. The procedure is illustrated with the Drude
model of metals.

Let ${\cal H}={\cal H}_{0}-i{\cal V}$ where ${\cal H}_{0}^{\dagger }={\cal H}%
_{0}$ and ${\cal V}^{\dagger }={\cal V}$. The system is absorbing and,
therefore, ${\cal V}$ must be a positive semidefinite operator, that is, for
any $\psi $, $(\psi ,{\cal V}\psi )\geq 0$. This readily follows from the
condition that the norm of a solution of (\ref{2.1}) cannot increase with
time. The exact time evolution (\ref{2.4}) is approximated by means of the
Lie-Trotter formula 
\begin{equation}
\psi (t+\Delta t)=e^{-\Delta t{\cal V}/2}e^{-i\Delta t{\cal H}%
_{0}}e^{-\Delta t{\cal V}/2}\psi (t)+O(\Delta t^{3})\ .  \label{3.1}
\end{equation}
The action of the exponential of ${\cal H}_{0}$ is computed by the Lanczos
method as before. The attenuation potential ${\cal V}$ typically does not
involve spatial derivatives and, hence, the action of its exponential on a
wave function is far less expensive than that for ${\cal H}_{0}$. The norm
of any power of the amplification matrix still remains uniformly bounded by
one because $\Vert e^{-\Delta t{\cal V}/2}\Vert \leq 1$ for $\Delta t\geq 0$%
. Hence the unconditional stability is preserved.

Let us turn to the Drude model which is used in numerical simulations
presented below. Another popular model, a multi-resonance Lorenz model, can
be treated similarly. Let ${\bf D}={\bf E}+{\bf P}$ and ${\bf B}={\bf H}$.
In the Drude model, the medium polarization is described by the second order
differential equation 
\begin{equation}
\ddot{{\bf P}}+\eta \dot{{\bf P}}=\omega _{p}^{2}{\bf E}\ ,  \label{3.2}
\end{equation}
where $\eta >0$ is the attenuation constant and $\omega _{p}$ is the plasma
frequency. Equation (\ref{3.2}) must be solved with zero initial conditions, 
${\bf P}(0)=\dot{{\bf P}}(0)=0$. Define an auxiliary field ${\bf Q}$ by $%
\dot{{\bf P}}=\omega _{p}{\bf Q}$. Rewriting the Maxwell's equations and (%
\ref{3.2}) in terms ${\bf E}$, ${\bf B}$, and ${\bf Q}$ and their
first-order time derivatives, the Schr\"{o}dinger equation is obtained in
which 
\begin{equation}
\psi =\pmatrix{{\bf E}\cr {\bf B}\cr {\bf Q}}\ ,\ \ \ \ {\cal H}=%
\pmatrix{0&ic\mbf{\nabla}\times & -i\omega_p\cr -ic\mbf{\nabla}\times
&0&0\cr i\omega_p&0&-i\eta }\ .  \label{3.3}
\end{equation}
The Hamiltonian is Hermitian when $\eta =0$ (no attenuation). The
attenuation potential ${\cal V}$ is a diagonal matrix, ${\rm diag}%
\,(0,0,\eta )$, that is positively semidefinite since $\eta >0$.

For an absorber at the grid boundaries, a layer of a conducting medium has
been used with a position dependent conductivity $\sigma $. As the induced
current in a conducting medium has the form ${\bf J}=$ $\sigma {\bf E}$, the
matrix ${\cal V}$ is changed to ${\rm diag}\,(-4\pi \sigma ,0,\eta )$. The
function $\sigma $ is constructed according to the frequency band of the
initial pulse.

\section{Free space propagation. Phase and amplitude errors}

\setcounter{equation}0

To illustrate the efficiency of the Lanczos time-propagation method, we
compare it with a widely adopted Second Order Finite Differencing (SOD, or
leapfrog) propagation method \cite{Review,Taf,Yee}, using the simplest
example of the electromagnetic pulse propagation in vacuum. The action of
the Hamiltonian on wave functions in the Lanczos and leapfrog methods are
done in the same way, that is, by the fast Fourier pseudospectral method on
the same grid.

Consider a Gaussian wave packet linearly polarized along the $y$ axis and
propagating along the $z$ axis. The amplitude of the fields at the initial
time $t=0$ is given by 
\begin{equation}
E_{y}(z)=e^{-z^{2}/D^{2}}e^{ik_{0}z}\ ,\ \ \ \ H_{x}(z)=-E_{y}(z)\ ,
\label{4.1}
\end{equation}
where $k_{0}=5.5/D$, and $D$ determines the width of the wave packet. The
carrier wave length $\lambda =2\pi /k_{0}$ so that $D=0.875\lambda $. We
take $D=1.75\,\mu m$, or $\lambda =2\,\mu m$. The step of the grid is $%
\Delta z=0.1D$. An exact solution directly follows from (\ref{4.1}) $%
E_{y}(z,t)=E_{y}(z-ct)$. The wave packet propagates in the direction of
positive $z$. With our settings the pulse duration is about $25\,fs$.

Numerical solutions are obtained by the Lanczos and leapfrog algorithms for
the Schr\"{o}\-dinger equation (\ref{2.1}) in which $\varepsilon =\mu =1$.
Recall that the leapfrog propagation scheme is based on the third-order
finite difference approximation of the time derivative 
\begin{equation}
\psi (t+\Delta t)=\psi (t-\Delta t)-2i\Delta t{\cal H}\psi (t)\ .
\label{4.2}
\end{equation}
The scheme is conditionally stable, and the time step must be chosen
accordingly. The simulated electric field is recorded by a detector placed
at $z=z_{\det }=18D$. Its phase and amplitude are compared with those of the
exact solution. For a signal $E(t)=E_{0}(t)e^{i\varphi (t)}$, where $%
E_{0}(t)=|E(t)|$, the phase and amplitude errors are defined, respectively,
by 
\begin{equation}
\delta ^{P}=\frac{|\varphi ^{exact}-\varphi ^{approx}|}{\varphi ^{exact}}\
,\ \ \ \ \delta ^{A}=\frac{|E_{0}^{exact}-E_{0}^{approx}|}{E_{0}^{exact}}\ .
\label{4.3}
\end{equation}
The errors $\delta ^{P,A}$ are plotted respectively in Figs 1 and 2 as
functions of $S=\left( z_{\det }-ct\right) /D$, the position of the pulse
center relative to the detector measured in units of $D$. The results are
shown for $\left| S\right| \leq 2.5$ where the signal on the detector
is sufficient. Dashed and solid lines correspond to the leapfrog and Lanczos
methods, respectively, for various settings of the time step.

The time step for the black dashed line is a reference time step, $\Delta
t_{0}\approx 0.01\,fs$. If $N_{H}$ is the number of elementary operations
required to compute the action of the Hamiltonian on a wave function, then
the total number of operations reads $N=sN_{H}N_{t}$, where $s$ is the
number of actions of the Hamiltonian per a time step, $N_{t}=t/\Delta t$ is
the total number of time steps. For the leapfrog method, $s=1$ for all time
steps. In the Lanczos method, $s=n-1$, with $n$ being the dimension of the
Krylov space. Despite that the dynamic control of accuracy has been
activated, {\it de facto} $n$ does not vary in due course of simulations in
vacuum.

Let $N=N_{0}$ for the black dashed curve. The red dashed curve is obtained
by reducing the time step, $\Delta t=\Delta t_{0}/2$, and, hence, the total
number of operations increases accordingly, $N=2N_{0}$. In the Lanczos
method, the black solid curve corresponds to $\Delta t=10\Delta t_{0}$ and $%
s=7$, the blue solid curve to $\Delta t=5\Delta t_{0}$ and $s=7$, and the
red one to $\Delta t=2.5\Delta t_{0}$ and $s=6$. The total number of
operations is, respectively, $N=0.7N_{0}$, $N=1.4N_{0}$, and $N=2.4N_{0}$.
It is readily seen that at roughly the same number of operations, the
Lanczos algorithm has phase and amplitude errors that are less than those in
the leapfrog method by several orders of magnitude.

A few remarks are in order. There are, of course, algorithms that would be
more efficient than the Lanczos propagation method in free space. For
instance, the split propagation method \cite{Split} essentially reproduces
an exact solution and is also unconditionally stable. However, the split
method would not be applicable when the Hamiltonian involves products of
operators that depend on spatial derivatives and positions. The accuracy of
the leapfrog scheme can be improved by, for example, taking into account the
next term of the Taylor expansion of $\psi (t\pm \Delta t)$ in powers of $%
\Delta t$ in (\ref{4.2}) \cite{Hagness}, 
\[
-2i\Delta t{\cal H}\psi \rightarrow -2i\Delta t{\cal H}(1-\Delta t^{2}{\cal H%
}^{2}/3)\psi \ . 
\]
In this case, $s=3$. The method is still conditionally stable where the
stability condition of the SOD, $\Delta t\Vert {\cal H}\Vert \leq 1$,
changes accordingly to $\Delta t\Vert {\cal H}(1-\Delta t^{2}{\cal H}%
^{2}/3)\Vert \leq 1$. Even though $s$ has tripled, the new stability
condition allows one to increase the time step by the factor of 2.1.
Therefore the total number of operations increases only slightly. The
accuracy of the scheme will be of $O(\Delta t^{5})$ which is still not as
high as in the Lanczos method, $O(\Delta t^{n})$ with $n=8,7$ in the above
examples.

Some care should be taken regarding a known drawback of the Lanczos
algorithm -- a possible loss of orthogonality of basis functions due to
round-off errors \cite{lanczos,OrtLoss}. This is why the time step has to be
adjusted so that only low dimensional Krylov spaces, $n\leq 9$, are invoked
in contrast to the conventional use of the Lanczos method for solving linear
systems.

\section{Applications to gratings}

\setcounter{equation}0

In this section the Lanczos propagation scheme is applied to the scattering
of a broad band wave packet on nanostructure periodic materials such as
gratings and grooves. We are particularly interested in transmission
(reflection) properties currently being a subject of intense research \cite
{PhotonicsGr,grVidal,MetGrat,Nanostructured}. The results obtained here are
compared with those available in the literature. The time-dependent approach
allows us to underline the role played by trapped modes or resonances in the
existence of extraordinary transmittance and reflectance of periodic
structures. The longer lives a trapped mode, the more narrow resonance
occurs in the reflection and/or transmission coefficient.

All systems considered here have a translation symmetry along one of the
Euclidean axes, chosen to be the $y$ axis. The structures are periodic along
the $x$ axis with the period $D_{g}$, and the $z$ direction is transverse to
the structure. The initial wave packet is Gaussian and propagates along the $%
z$ axis. Its spectrum is broad enough to cover the frequency range of
interest. The zero diffraction mode is studied for wavelengths $\lambda
\geq D_{g}$ so that reflected and transmitted beams propagate along the 
$z$-axis. As in our previous work \cite{bs} we use a change of variables to
enhance the sampling efficiency in the vicinity of medium interfaces so that
the boundary conditions at sharp interfaces are accurately reproduced by the
Fourier-grid pseudospectral method. A typical size of the mesh corresponds
to $-15D_{g}\leq z\leq 15D_{g}$, and $-0.5D_{g}\leq
x\leq 0.5D_{g}$ with, respectively, $512$ and $128$ knots. The
frequency resolved transmission and reflection coefficients are obtained via
the time-to-frequency Fourier transform of the signal on ``virtual
detectors'' placed at some distance in front and behind the slab with a
periodic structure \cite{VirtualDet}.

\subsection{Array of dielectric cylinders}

The significance of trapped modes is first illustrated with a periodic array
of non-dispersive dielectric cylinders, the system which has not received as
much attention as metal or dielectric gratings. Consider an array of
parallel, periodically positioned, dielectric cylinders in vacuum oriented
along the $y$ axis. The radius $R$ of cylinders is small as compared to the
array period $D_{g}=1.75\mu m$. In simulations, the ratio $R/D_{g}$ is taken
to be $0.0857$. The incident wave packet is linearly polarized. The electric
field is oriented along the $y$ axis, i.e., parallel to the cylinders (the
so called TM polarization). The Hamiltonian for the Lanczos scheme has the
form (\ref{2.1}) where $\mu =1$.

In Fig. 3 the reflection coefficient ${\cal R}$ is shown as a function of
the wave length expressed in units of $D_{g}$. In the Schr\"{o}dinger
formulation of Maxwell's theory the norm of the wave function is
proportional to the total electromagnetic energy. Hence, for a lossless
medium the transmission ${\cal T}$ can simply be obtained from the energy
conservation: ${\cal T}+{\cal R}=1$. Recall that the Lanczos propagation
method preserves the norm. The solid-blue and dashed-red curves correspond,
respectively, to $\varepsilon =2$ and $\varepsilon =4$. As one can see the
array becomes a perfect reflector within a fairly narrow wavelength range
centered at the resonant wavelength that is slightly larger than the period $%
D_{g}$. Similar results have been obtained for dielectric grating
structures. The resonant pattern is associated with the so-called Wood
anomalies \cite{Wood}, and can be explained by the existence of trapped
modes or guided wave resonances \cite{bs,DielGratings}. The widths of the
resonances in the reflection (transmission) coefficient are determined by
the lifetime of a corresponding quasi-stationary trapped mode which is a
standing wave along the $x$ axis and is excited by the incoming wave.

The existence of trapped modes can easily be inferred from the temporal
evolution of the electromagnetic field. Figure 4 shows the transmitted
electric field as a function of time measured by a detector placed behind
the layer of dielectric cylinders. The main transmitted pulse is clearly
visible. It has a significant amplitude and duration about $25\,fs$. After
the main pulse passes the array, it leaves behind an excited
quasi-stationary mode which looses its energy by radiating almost
monochromatic waves with the same amplitude, but an opposite phase, in the
transmission and reflection directions. The lasing effect of the trapped
mode appears as exponentially dumped oscillations coming after the main
signal. The exponential decay due to a finite lifetime of the
quasi-stationary state is clearly seen. By the symmetry, the same lasing
effect is registered by a detector placed in front of the layer (not shown
here). A 100\% reflection at the resonant frequency can be understood from
the fact that the field emitted by the trapped mode in the transmission
direction and the corresponding frequency component of the initially
transmitted pulse have an opposite phase, thus compensating each other. The
solid-blue and dashed-red curves correspond, respectively, to $\varepsilon
=2 $ and $\varepsilon =4$. The radiation coming from the narrow resonance
(the blue curve) has a lower amplitude and a much longer duration. The
lifetime of the trapped mode in this case is in the picosecond range, i.e.,
thousand times longer than the initial pulse duration. Note that the more
narrow resonance is the less energy gets trapped from the initial pulse.
This explains the amplitude difference of the blue and red curves. Finally,
the concept of trapped modes localized on successive layers and interacting
with each other provides a theoretical framework for the light propagation
in layered structures such as photonic crystal slabs \cite{Slabs}.

\subsection{Metal gratings and grooves}

Metal gratings and grooves have been extensively studied in micro-wave and
optical domains \cite{grVidal,MetGrat}. The purpose of this section is to
show that the Lanczos propagation method can successfully be applied to
metals described by the Drude model. The Hamiltonian has the form (\ref{3.3}%
). The attenuation and the plasma frequency are taken to be representative
for silver: $\omega _{p}=9eV$, and $\eta =0.1eV$ \cite{grVidal}. The grating
geometry is sketched in the inset of Fig. 5. The grating period is $%
D_{g}=1.75\,\mu m$, the thickness (along the $z$ axis) is $h=0.8\,\mu m$,
and the grating width $a=0.3\,\mu m$. The corresponding grooves are obtained
by attaching a solid metal plate on one side of the gratings so that no
transmission is possible. The polarization of the incident wave packet is
such that the electric field vector is oriented along the $x$ axis, i.e.,
perpendicular to the gratings (the so called TE polarization). The
difference with the non-dispersive case discussed above is the presence of
attenuation. The trapped mode looses its energy due to (non-perfect)
conductivity of the metal. This leads to broadening of the resonance.

In Fig.5 the dashed red and solid blue curves represent the transmission and
reflection coefficients, respectively, as functions of the wavelength
expressed in units of the grating period, $D_{g}$. The resonance is again
associated with the existence of a trapped stationary wave in the grating.
The transmittance does not reach 100\% due to dissipative loss of energy in
the Drude metal. While for a lossless medium the sum of the reflection and
transmission coefficients must be one, this is not the case for the Drude
metal (the dashed-dotted green curve in Fig. 5). The maximal loss of energy
corresponds to the resonant wavelength. It is easily understood because the
trapped mode remains in contact with the metal much longer than the main
pulse, and, therefore, can dissipate more energy through exciting surface
electrical currents in metal. The black curve in Fig. 5 shows the
reflectance of the grooves. Since the light cannot be transmitted through
the grooves, a resonance structure in the reflection coefficient is directly
related to the enhanced energy loss at the wavelength of the trapped mode.
Note that as compared to the metal gratings, the resonance is broadened and
shifted to the lower frequencies (larger wavelength). The results obtained
here are in a full agreement with previous theoretical and numerical
analysis \cite{grVidal,MetGrat}.

\section{Conclusions}

\setcounter{equation}0

It has been demonstrated that the Lanczos algorithm can be used to develop a
highly efficient, accurate, and unconditionally stable propagation scheme to
simulate scattering of broad band electromagnetic pulses in passive media.
The accuracy and efficiency of the algorithm have been illustrated with an
example of the electromagnetic wave propagation in vacuum. At the same
computational costs, a significant reduction of phase and amplitude errors
has been observed in the Lanczos propagation method as compared to the
second-order finite-difference (leapfrog) scheme.

As an example of possible applications, the Lanczos propagation method has
been applied to study resonant transmission and reflection of various
periodic nano-structures: An array of periodically placed parallel cylinders
made of a non-dispersive dielectric material, metallic gratings and grooves.
The time-domain study clearly demonstrates the role played by
quasi-stationary (trapped) electromagnetic waves supported by the
corresponding periodic structure in the extraordinary transmission
(reflection) properties of the grating. The results for metallic gratings
and grooves coincide with those obtained earlier by means of other numerical
algorithms and are also in agreement with theoretical studies. The
unconditional stability of the Lanczos propagation scheme for media with
attenuation has been achieved via the split method, which reduces the
accuracy. It is possible to restore the accuracy up to the level gained for
non-absorbing media. However, stability conditions require a further study
that will be reported elsewhere.

In summary, the Lanczos algorithm has been shown to lead to a highly
accurate, efficient, and unconditionally stable time-propagation numerical
solver for the Maxwell's equations. Variable time steps and/or variable
computational costs with accuracy control are possible. The method is
applicable to various electromagnetic systems (no restrictions on the
Hamiltonian). All these virtues are hardly available in other
unconditionally stable algorithms in numerical electrodynamics of passive
media.

\vskip 1cm {\bf Acknowledgments}

We acknowledge the financial support and hospitality of Donostia
International Physics Center (DIPC). S.V.S. thanks the director of LCAM, Dr.
V. Sidis, for his continued support and kind hospitality.
S.V.S. is also grateful to Dr. R. Albanese
(US Air Force Brooks Research Center, TX), Profs. J.R. Klauder and T. Olson
(University of Florida) for the support of this project.

\newpage
{\bf Figure captions}

Fig. 1.\qquad The phase errors for the propagation of an electromagnetic
gaussian pulse in vacuum. Results are presented as a function of the
position of the pulse center relative to the detector, $S$, measured in
units of the pulse width $D$. Dashed and solid curves correspond,
respectively, to the leapfrog and Lanczos propagation methods. Different
colors represent computational costs of simulations measured as the total
number of actions of the Hamiltonian on the wave function for fixed
propagation time. Further details are given in the text.\bigskip

Fig. 2.\qquad The amplitude errors for the propagation of an electromagnetic
gaussian pulse in vacuum. Results are presented as a function of the
position of the pulse center relative to the detector, $S$, measured in
units of the pulse width $D$. Dashed and solid curves correspond,
respectively, to the leapfrog and Lanczos propagation methods. Different
colors represent computational costs of simulations measured as the total
number of actions of the Hamiltonian on the wave function for fixed
propagation time. Further details are given in the text.\bigskip

Fig. 3\qquad Calculated zero-order reflection coefficient for a periodic
array of dielectric cylinders in vacuum described in the text. Results are
presented as a function of the wavelength of the incident radiation measured
in units of the period $D_{g}$. The solid blue and dashed red curves
correspond, respectively, to the array of cylinders with dielectric
constants $\varepsilon =2$ and $\varepsilon =4$. \bigskip

Fig. 4\qquad The electric field measured by a detector placed behind the
periodic layer of dielectric cylinders. Only the field corresponding to the
zero-order transmitted wave propagating along the $z$-axis is represented.
It is obtained by the Fourier analysis of the $x$-coordinate dependence of
the field at the detector position. The signal is shown as a function of
time measured in femtoseconds. The solid blue and dashed red curves
correspond, respectively, to the array of cylinders with dielectric
constants $\varepsilon =2$ and $\varepsilon =4$. \bigskip

Fig. 5.\qquad Calculated zero-order reflection and transmission coefficients
for metallic gratings and groves described in the text. Results are
presented as a function of the wavelength of the incident radiation measured
in units of the period $D_{g}$. The inset of the figure gives a schematic
view on the grating geometry. The black line shows the reflection
coefficient for metallic grooves. Blue and (dashed red) line shows the
reflection (transmission) coefficient for metallic gratings. The sum of the
reflection and transmission coefficients for metallic gratings is shown as
the dashed-dotted green curve. Its deviation from $1$ represents the loss of
electromagnetic energy because of the absorption in metal.

\end{document}